\documentclass[a4paper,11pt]{article} 
\usepackage[margin=0.5in]{geometry} 
\usepackage{amssymb}
\usepackage{amsmath,epsfig, epsf, epic, graphicx}
\usepackage{xcolor}
\usepackage{fancybox, lscape, subfigure} 
\usepackage{bm}
\usepackage{natbib}
\usepackage{float}
\usepackage[margin=2cm]{caption}
\usepackage{setspace}

\newcommand\independent{\protect\mathpalette{\protect\independenT}{\perp}}
\def\independenT#1#2{\mathrel{\rlap{$#1#2$}\mkern2mu{#1#2}}}

\title{Averaging causal estimators in high dimensions}
\author{Joseph Antonelli and Matthew Cefalu}

\allowdisplaybreaks

\begin{document}
\date{}
\maketitle{}

\begin{abstract}
    There has been increasing interest in recent years in the development of approaches to estimate causal effects when the number of potential confounders is prohibitively large. This growth in interest has led to a number of potential estimators one could use in this setting. Each of these estimators has different operating characteristics, and it is unlikely that one estimator will outperform all others across all possible scenarios. Coupling this with the fact that an analyst can never know which approach is best for their particular data, we propose a synthetic estimator that averages over a set of candidate estimators. Averaging is widely used in statistics for problems such as prediction, where there are many possible models, and averaging can improve performance and increase robustness to using incorrect models. We show that these ideas carry over into the estimation of causal effects in high-dimensional scenarios. We show theoretically that averaging provides robustness against choosing a bad model, and show empirically via simulation that the averaging estimator performs quite well, and in most cases nearly as well as the best among all possible candidate estimators. Finally, we illustrate these ideas in an environmental wide association study and see that averaging provides the largest benefit in the more difficult scenarios that have large numbers of confounders. 
\end{abstract}

\section{Introduction}

Estimating the causal effect of a treatment on an outcome with observational data relies on the assumption that the treatment assignment is unconfounded conditional on a set of pre-treatment covariates. There is a large literature on methods to adjust for confounding in order to satisfy this assumption, such as matching \citep{rubin1996matching}, propensity score weighting \citep{rosenbaum1983central}, stratification \citep{rosenbaum1984reducing}, outcome regression, and doubly robust estimators \citep{bang2005doubly}. These approaches have all been shown to provide valid estimates of causal effects as long as the covariates necessary for the unconfoundedness assumption to hold are measured in the data. In recent years, the estimation of causal effects in high-dimensional data sets with large numbers of covariates have become increasingly common. A large number of measured covariates makes it more plausible that the covariates necessary for confounding adjustment have been collected; however, estimation becomes considerably more difficult with large numbers of parameters. Due to this trade off, methods have been developed to adjust for large numbers of covariates and produce consistent estimates of causal effects from observational data.

Much of the interest began as papers started to consider the problem of variable selection or model selection for causal estimation in low-dimensional settings \citep{crainiceanu2008adjustment,vansteelandt2012model, wang2012bayesian}. Since then, a large number of manuscripts have been published that aim to estimate causal effects when the number of covariates is large relative to the sample size. \cite{wilson2014confounder} tailored the adaptive lasso \citep{zou2006adaptive} to effect estimation instead of prediction, by creating weights that involve both outcome model and treatment model coefficients. Other approaches have utilized penalization approaches to estimate causal effects in high-dimensions. \cite{shortreed2017outcome} used the adaptive lasso to estimate the propensity score, but allowed the weights to depend on the outcome model coefficients. \cite{ertefaie2018variable} created a new penalized likelihood estimator that utilizes information from both the treatment and outcome to identify confounders, and then estimate causal effects conditional on this set of variables. The lasso \citep{tibshirani1996regression} was used to estimate both propensity score and prognostic score models in \cite{antonelli2016double} to obtain doubly robust estimates by matching on both estimated scores. \cite{antonelli2017high} utilized a fully Bayesian approach to estimating treatment effects in high-dimensions that reduces finite sample bias while eliminating the impact of instrumental variables through informative prior distributions on variable inclusion parameters. \cite{hahn2016bayesian} utilized horseshoe priors on the coefficients from a re-parameterized likelihood to reduce shrinkage of variables that are associated with both the treatment and outcome. A large number of approaches have focused on obtaining uniformly valid inference of causal effects in high-dimensions \citep{belloni2013inference,farrell2015robust,chernozhukov2018double}. \cite{belloni2013inference} used lasso models to perform variable selection on both the treament and outcome models, taking the union of the identified covariates from the two models, and then estimating effects conditional on the chosen model. \cite{farrell2015robust} also utilized the lasso for a treatment and outcome model selection, and then fit post selection estimators to estimate the traditional doubly robust estimator. \cite{chernozhukov2018double} estimated nuisance parameters such as treatment or outcome model parameters via penalization, and then solved the efficient score to estimate treatment effects. Similar ideas are used from the Bayesian perspective in \cite{antonelli2018bayesian}, and it is shown that it leads to improved finite sample performance of frequentist metrics such as interval coverage. Other approaches have shown that $\sqrt{n}$ convergence of treatment effects can be achieved in high-dimensional settings. \cite{athey2018approximate} combined high-dimensional outcome regression with balancing weights that eliminate any residual confounding remaining after the outcome regression. Propensity score estimation is improved for both high-dimensions and model misspecification using calibration weights in \cite{tan2017regularized, tan2018model}. Propensity scores were again used in high-dimensional scenarios, but are instead estimated directly by their ability to balance important covariates in \cite{ning2018robust}. Lastly, targeted maximum likelihood (TMLE, \cite{van2006targeted}) can be used in conjuction with high-dimensional models to produce estimates of treatment effects. Recent work has tailored TMLE to this scenario using collaborative TMLE (\cite{ju2019scalable, ju2019collaborative}) or variance moderation (\cite{hejazi2017variance}).

These developments have led to a diverse and large set of estimators to use if one is faced with observational data that is high-dimensional. Choosing among this large set of estimators is a subjective choice, and it is not clear if, or when, any estimators are preferable to others. Given the complexity faced with high-dimensional data, it is highly unlikely that any one estimator will perform well universally. 

In this manuscript, we leverage a large set of potential estimators to our advantage by introducing a simple, intuitive estimator that averages over all of the candidate estimators. We show both theoretically and empirically that this leads to robust performance for estimating treatment effects in high-dimensions. Through extensive simulation studies we show that averaging over all estimators generally leads to improved performance with respect to standard metrics such as bias, mean squared error (MSE), and coverage probabilities. We illustrate through a wide variety of data generating mechanisms that there is not any one individual estimator that performs well across all scenarios; however, the proposed averaged estimator performs nearly as well as the best estimator in the majority of cases considered, which shows the robustness that averaging provides for high-dimensional treatment effect estimation. 

The idea of combining models, estimators, or tests has a rich history in the statistical literature. For example, model uncertainty can be accounted for through model averaging, which exists in both the Bayesian and frequentist framework (\cite{draper1995assessment,hoeting1999bayesian, hjort2003frequentist, hansen2007least}). Others have proposed combinations of estimators (\cite{mittelhammer2005combining, antonelli2017synthetic}), including in the field of small area estimation (\cite{ghosh1994small}). Omnibus tests for the global null hypothesis simultaneously test multiple hypotheses (e.g. ANOVA), and several methods for combining individual p-values from genome-wide association studies have been proposed (\cite{bbx068}). These are just a few examples of combining models, estimators, or tests, but they all share a common goal: combining statistics for improved performance. 

\section{Methodology}

Throughout, we assume that we have $n$ i.i.d samples of $(\boldsymbol{X}_i, T_i, Y_i)$, where $\boldsymbol{X}_i$ is a $p-$dimensional vector of pre-treatment covariates, $T_i$ is the treatment of interest, and $Y_i$ is the outcome of interest that is measured after the treatment is initiated. We assume that $T_i$ is a binary treatment as many of the existing estimators are only applicable to binary treatments.  We do not discuss the rates at which $p$ grows, but work in the high-dimensional regime where $p > n$ and grows with $n$. For simplicity, we restrict attention to the average treatment effect defined as $\text{ATE} = E(Y(1) - Y(0))$, where $Y_i(t)$ is the potential outcome we would have observed for subject $i$ if they had been exposed to treatment $T_i = t$. We assume no unmeasured confounding and positivity, defined below: \\

\indent \textit{Unconfoundedness:} $Y(t) \independent T \vert \boldsymbol{X}$ for t=0,1 \\
\indent \textit{Positivity:} There exist $\delta \in (0,1)$ such that $0 < \delta < P(T=1 \vert \boldsymbol{X}) < 1-\delta < 1$ with probability 1,
\\
\\
\noindent where $P(T=1 \vert \boldsymbol{X})$ denotes the propensity score \citep{rosenbaum1983central}. We also assume the stable unit treatment value assumption (SUTVA), which states that the treatment is well-defined and that the outcomes for one subject are not affected by the treatment status of other subjects. Further, in high-dimensional settings, many estimators rely on some form of sparsity, stating that unconfoundedness holds conditional of a subset of the covariates. The exact form of the sparsity condition varies by estimator and is not discussed in detail in this manuscript.

\subsection{Averaged estimator}

Consider constructing an averaged estimator from a set of $M$ estimators for the ATE. Denote each estimator of the ATE by $\widehat{\theta}_m$ for $m=1,\dots, M$. Each estimator has a corresponding variance estimate, denoted by $\widehat{\sigma}_m^2$ for $m=1,\dots,M$. The averaged estimator takes the form:

$$\widehat{\theta}_A = \sum_{m=1}^M w_m \widehat{\theta}_m,$$

\noindent where $\boldsymbol{w}=(w_1,\dots,w_M)$ is a weight vector. Ideally, we 
would estimate $\boldsymbol{w}$ by optimizing a pre-specified loss function. Following \cite{antonelli2017synthetic}, one could find the weight vector that minimizes the mean squared error of $\widehat{\theta}_A$ conditional on $\boldsymbol{w}$, which corresponds to solving:

$$\underset{\boldsymbol{w}}{\operatorname{arg min}} \ \boldsymbol{w}^T \boldsymbol{V} \boldsymbol{w} + (\boldsymbol{w}^T\boldsymbol{B})^2,$$

\noindent where $\boldsymbol{B}$ and $\boldsymbol{V}$ are the bias vector and covariance matrix for the $M$ estimators. Finding the weight vector that minimizes the mean squared error requires estimates of both $\boldsymbol{B}$ and $\boldsymbol{V}$. In the context of noncompliance in clinical trials, \cite{antonelli2017synthetic} were able to estimate these quantities. The bias was estimated by identifying an unbiased estimator among the candidate estimators, and the covariance was estimated using the bootstrap. In the case of observational studies, the bias of each estimator is unknown a priori and there do not exist methods for estimating it from observed data. 

The covariance matrix $\boldsymbol{V}$ of the estimators is also problematic when there are high-dimensional confounders. The standard bootstrap is generally not valid for estimating sampling distributions of high-dimensional models such as the LASSO \citep{chatterjee2010asymptotic}. In the context of high-dimensional ATE estimation, it has been seen empirically that the bootstrap can lead to poor inferential properties such as incorrect standard error estimation \citep{antonelli2018bayesian}. Due to this, much of the existing work on estimating the ATE in high dimensions has focused on deriving asymptotic limiting distributions for each estimator. While these asymptotic results may provide variance estimates for individual estimators, estimation of $\boldsymbol{V}$ also requires the pairwise covariance between each estimator. It is extremely cumbersome, and perhaps not possible, to derive the covariance for each pair of estimators, and would require additional work whenever new candidate estimators are included.  Due to the difficulty of estimating $\boldsymbol{B}$ and $\boldsymbol{V}$, we will proceed forward with a fixed weight vector $\boldsymbol{w} = (1/M, 1/M, \dots, 1/M)$, such that each estimator gets equal weight. This provides a number of advantages over simply choosing a single estimator, which are detailed in the subsequent sections. The averaging provides a form of robustness ensuring that the averaged estimator does better in terms of bias and MSE than the worst of the $M$ estimators. The empirical studies in Section \ref{sec:sim} show that averaging not only provides protection against choosing the worst estimator, but the averaged estimator typically does close to, if not better, than the best of the $M$ estimators. This is due to large decreases in the variance of the averaged estimator relative to any one estimator, particularly if the individual estimators are not strongly correlated. Note that unequal weights can easily be applied as long as they are chosen a priori and not chosen simply to provide a desired result. Further, if there is concern about one bad estimator, in particular one with large amounts of bias, then a trimmed average can be used. We explore this further in the simulation study of section \ref{sec:sim} where we use a trimmed average estimator that removes the highest and lowest estimates from the averaged estimator. 

\subsection{Variance estimation}
\label{sec:var}

If the covariance matrix $\boldsymbol{V}$ of the $M$ estimators is known, the variance of the averaged estimator can be written as:

$$\text{var}(\widehat{\theta}_A) = \frac{1}{M^2} \sum_{i=1}^M \sum_{j=1}^M V_{ij}.$$

\noindent Exploring this quantity provides insight in how the averaged estimator can greatly reduce variance. In one extreme case where all the estimators are perfectly correlated, then the variance of the averaged estimator is equal to the average of the pairwise products of the standard deviations for each estimator, implying that its variance is no larger than the highest variance estimator. In a different extreme case where all of the estimators are uncorrelated, the variance of the averaged estimator is $\frac{1}{M^2} \sum_{i=1}^M V_{ii}$. This implies that the variance of the averaged estimator is an order of magnitude ($1/M$) smaller than the average of the individual estimator variances. Even bigger gains are possible if the estimators are negatively correlated, but this is very unlikely to happen in practice. In our experience, most estimators are positively, but not perfectly, correlated with each other, meaning that the truth lies between the two extremes discussed. Nonetheless, the variance of the averaged estimator will be no worse than the highest variance estimator, and may be a substantial improvement over the individual estimators. 

Estimating the variance of the averaged estimator is problematic due to the underlying high-dimensional estimators. A plug-in estimate of the variance based on the expression above is not feasible, because it is very difficult to produce the necessary covariances in high-dimensions. We can, however, estimate the individual estimator variances $\widehat{\sigma}_m^2$. Since the correlation between estimators is bounded between -1 and 1, we can bound the variance of the averaged estimator as:

$$\text{var}(\widehat{\theta}_A) = \frac{1}{M^2} \sum_{i=1}^M \sum_{j=1}^M V_{ij} = \frac{1}{M^2} \sum_{i=1}^M \sum_{j=1}^M \text{cor}(\widehat{\theta}_i, \widehat{\theta}_j) \sqrt{V_{ii}} \sqrt{V_{jj}} \leq \frac{1}{M^2} \sum_{i=1}^M \sum_{j=1}^M \sqrt{V_{ii}} \sqrt{V_{jj}}.$$

\noindent Therefore, we propose a conservative estimate of the variance as:

$$\widehat{\sigma}_A^2 = \frac{1}{M^2} \sum_{i=1}^M \sum_{j=1}^M \widehat{\sigma}_i \widehat{\sigma}_j.$$

\noindent If the estimators are highly, positively correlated, this variance estimate will not be very conservative. If the estimators are close to independent, then this variance estimator is very conservative. It is possible that being conservative in this setting is beneficial for two reasons. First, the estimates of standard errors $\widehat{\sigma}_m$ frequently rely on asymptotic approximations, which might not hold, and could be underestimates of the true variances. Second, there is frequently bias in estimating treatment effects in high-dimensions due to penalization or variable selection, and conservative estimates of the variance may lead to better interval coverage. Section \ref{sec:sim} provides empirical evidence of the performance of this variance estimation strategy.

\subsection{Robustness to choosing the worst model}

The relative performance of the averaged estimator to the candidate estimators is highly data dependent. However, we show that the averaged estimator protects against choosing a very bad estimator among the class of estimators. Intuitively, this is because the averaged estimator only assigns a weight of $1/M$ to any one estimator, including the worst estimator with respect to a given metric; therefore, the averaged estimator is guaranteed to perform no worse than this worst case estimator. The following result shows this for both bias and mean squared error: \\
\\
\textit{Result 1:} If each estimator is assigned equal weight, and we let $\widehat{\theta}^*$ and $\widehat{\theta}^{**}$ be the worst estimators in terms of bias and MSE, respectively, then

$$|E(\widehat{\theta}_A - \theta)| \leq |E(\widehat{\theta}^* - \theta)| \text{ and } E((\widehat{\theta}_A - \theta)^2) \leq E((\widehat{\theta}^{**} - \theta)^2)$$

\noindent A proof of this result can be found in the Appendix. While this result only guarantees that the averaged estimator does better than the worst estimator, empirically we have seen that these are typically very conservative bounds. In simulation studies in Section \ref{sec:sim} we illustrate that the averaged estimator not only performs better than the worst estimators, but frequently performs as well as the best individual estimator in terms of MSE. \\

\section{Simulation study}
\label{sec:sim}

We empirically investigate the ability of the averaged estimator to estimate treatment effects in high-dimensional settings. We focus on 10 potential estimators that are either available in an R package, or straightforward to implement. An R package to implement these estimators along with the averaged estimator can be found at \url{https://github.com/jantonelli111/AveragingCausalHD}. Computation times for the estimators can also be found in Section B of the Appendix. The estimators that enter into the averaged estimator are listed below:

\begin{enumerate}
    \item The double post selection approach (Double PS, \cite{belloni2013inference})
    \item The approximate residual debiasing approach (Debiasing, \cite{athey2018approximate})
    \item The doubly robust lasso approach (DR-Lasso, \cite{farrell2015robust})
    \item Doubly robust matching (DRME, \cite{antonelli2016double})
    \item High-dimensional confounding with spike-and-slab priors (HDC, \cite{antonelli2017high})
    \item Bayesian doubly robust estimator (DR-Bayes, \cite{antonelli2018bayesian})
    \item Targeted maximum likelihood with lasso models (TMLE, \cite{van2006targeted})
    \item Targeted maximum likelihood with an initial screening step followed by unpenalized linear models (TMLE-screen)
    \item Double machine learning with lasso models (DML, \cite{chernozhukov2018double})
    \item Double machine learning with post selection estimators (DML-PS)
\end{enumerate}

\noindent We do not go into the details of the respective approaches or how they are fit, as the aim of this work is to show the performance of the averaged estimator when there are a large number of estimators to average over, each with varying operating characteristics that are unknown to the user beforehand. For this reason we will only highlight the best, median, and worst estimator among the individual estimators with respect to any particular metric. We will also show results for two averaged estimators: The averaged estimator with equal weights for each estimator, and a trimmed average that removes the estimators with the highest and lowest estimated causal effects within a particular data set. For mean squared error, the best estimator is the one with the smallest MSE, the worst is the one with the largest MSE, and the median is the median MSE among the individual estimators. For ease of discussion, the MSE for each simulation scenario is normalized so that the best estimator has MSE equal to 1. For interval coverage, the worst estimator is the one with the lowest coverage, and then the best estimator is the one with the coverage closest to 0.95 without being 1 We want to avoid an estimator with a coverage probability of 1 as it is overly conservative. Again the median estimator is simply the median coverage probability among the individual estimators. 

The exact details of the simulation scenarios can be found in Section C of the Appendix, though we will briefly summarize them here. In each simulation scenario we draw covariates from a multivariate normal distribution with correlation 0.3 between all covariates. We will set $n=150$ and $p=300$ for all simulation scenarios with the exception of scenario 2, which will have $n=300$ subjects. The first simulation scenario represents a data generating mechanism that has a sparse, linear model for both the treatment and outcome. The second simulation scenario is one in which the treatment model is dense and all covariates are associated with the treatment. The third scenario looks at a setting where the outcome model is not sparse. The fourth setting is one in which the outcome model is nonlinear and therefore many of the approaches will be misspecified. The final simulation setting is one in which there is treatment effect heterogeneity, which also can cause misspecification of some of the approaches that assume homogeneous effects. These are chosen to capture a wide range of data generating mechanisms that might occur in practice. We simulated five additional scenarios, and the results can be found in Section C of the Appendix. The results in the additional simulation scenarios closely match those seen in the main manuscript.

\subsection{Results}

\begin{figure}[H]
\centering
	  \includegraphics[width=0.9\linewidth]{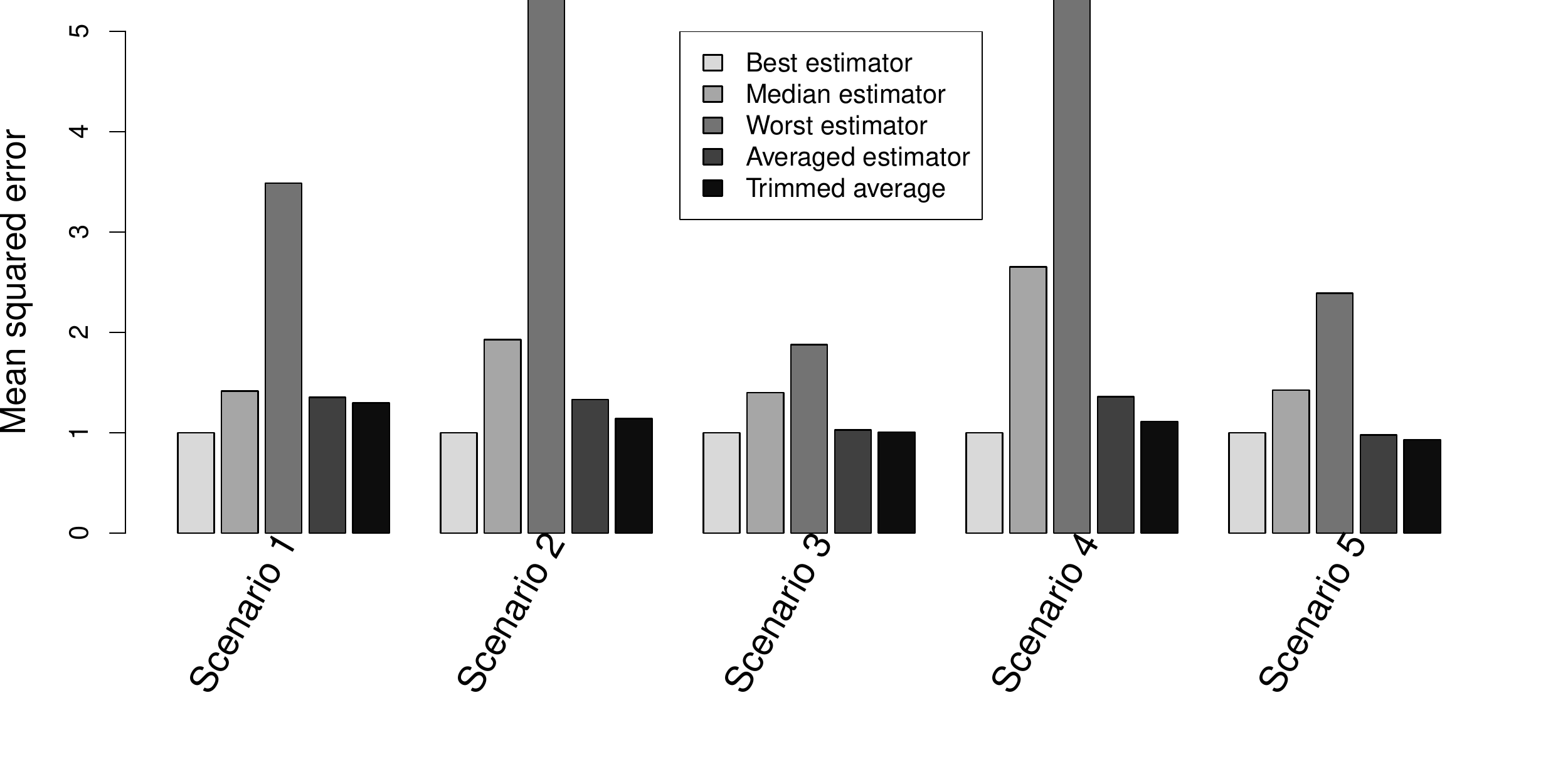}
\caption{Mean squared error for the estimator with the lowest MSE, highest MSE, median MSE, the averaged estimator, and the trimmed average estimator.}
\label{fig:simMSE}
\end{figure}

\begin{figure}[H]
\centering
	  \includegraphics[width=0.9\linewidth]{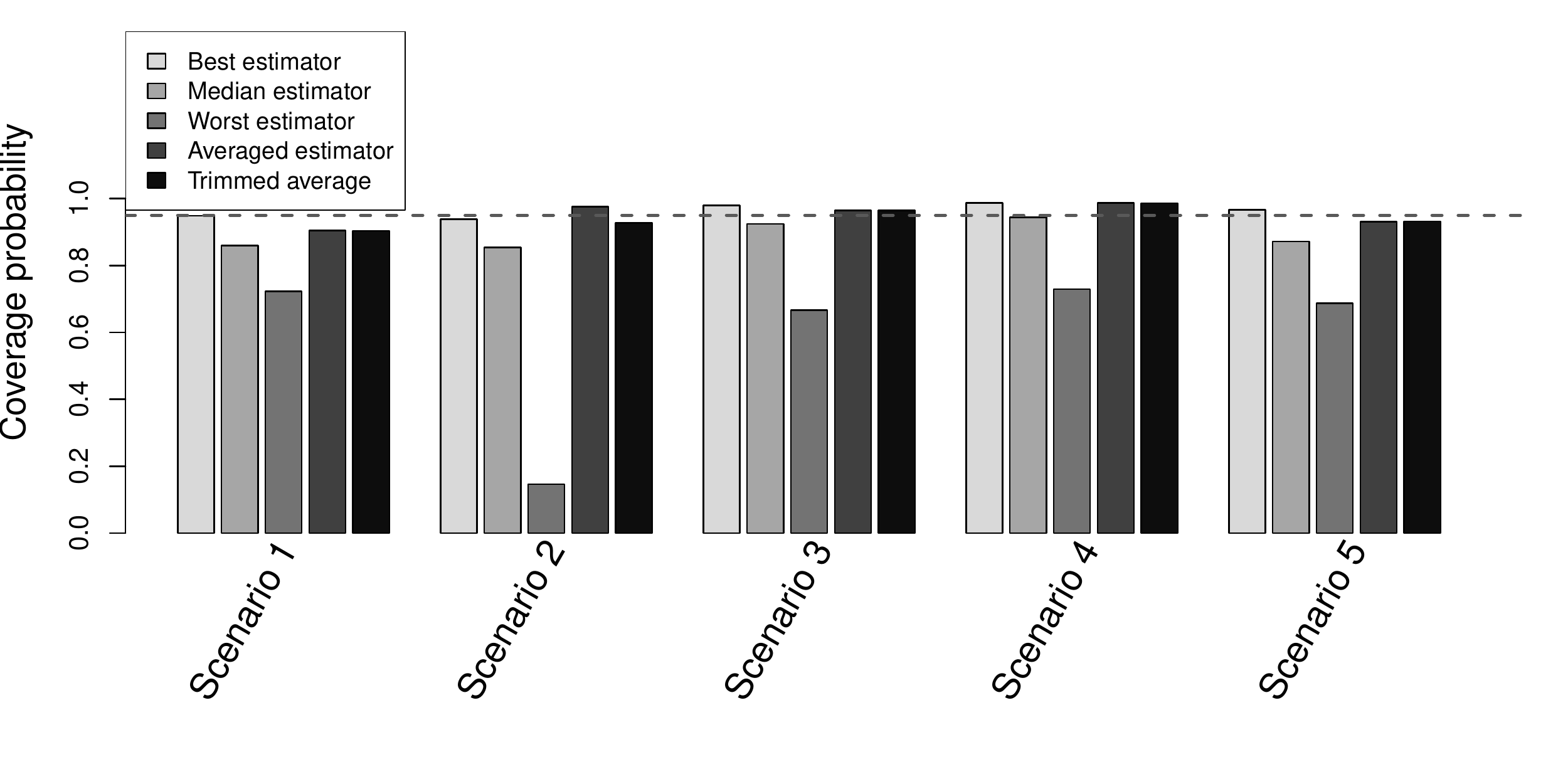}
\caption{Interval coverage for the estimator with the best interval coverage, median interval coverage, worst interval coverage, the averaged estimator, and for the trimmed average estimator.}
\label{fig:simINT}
\end{figure}

Figures \ref{fig:simMSE} and \ref{fig:simINT} show the results averaged over 1000 simulations for both MSE and interval coverage. Both averaging estimators performed favorably in terms of MSE and interval coverage. The highest MSE estimator in each scenario had significantly higher MSE than the averaging estimators. The averaging estimators had lower MSE than the median MSE estimator in every scenario and were able to achieve MSE that is close to the optimal choice of the individual estimators. They had slightly lower MSE than the individual estimator with the lowest MSE in Scenario 5. The averaging estimators achieved confidence interval coverage near the nominal 0.95 level in all of the simulation scenarios, and significantly outperformed both the worst estimator and the median estimator. 

Overall, the results confirmed the theoretical results that the averaged estimator does better than the worst individual estimator, though more importantly, the simulation showed that averaging estimators tends to perform more closely to the best individual estimator (at least in the scenarios considered). It should be noted that the individual estimator that performed the best or worst in each scenario was different, and therefore there is not one estimator outperforming the averaged estimator across all scenarios. Further, the results showed that the conservative estimate of the variance is not leading to overly conservative confidence intervals for the averaging estimators. The trimmed average estimator did not provide drastically different results from the averaging estimator that assigns equal weights to all of the estimators, though it did tend to reduce the MSE slightly.

\begin{figure}[H]
\centering
	  \includegraphics[width=0.45\linewidth]{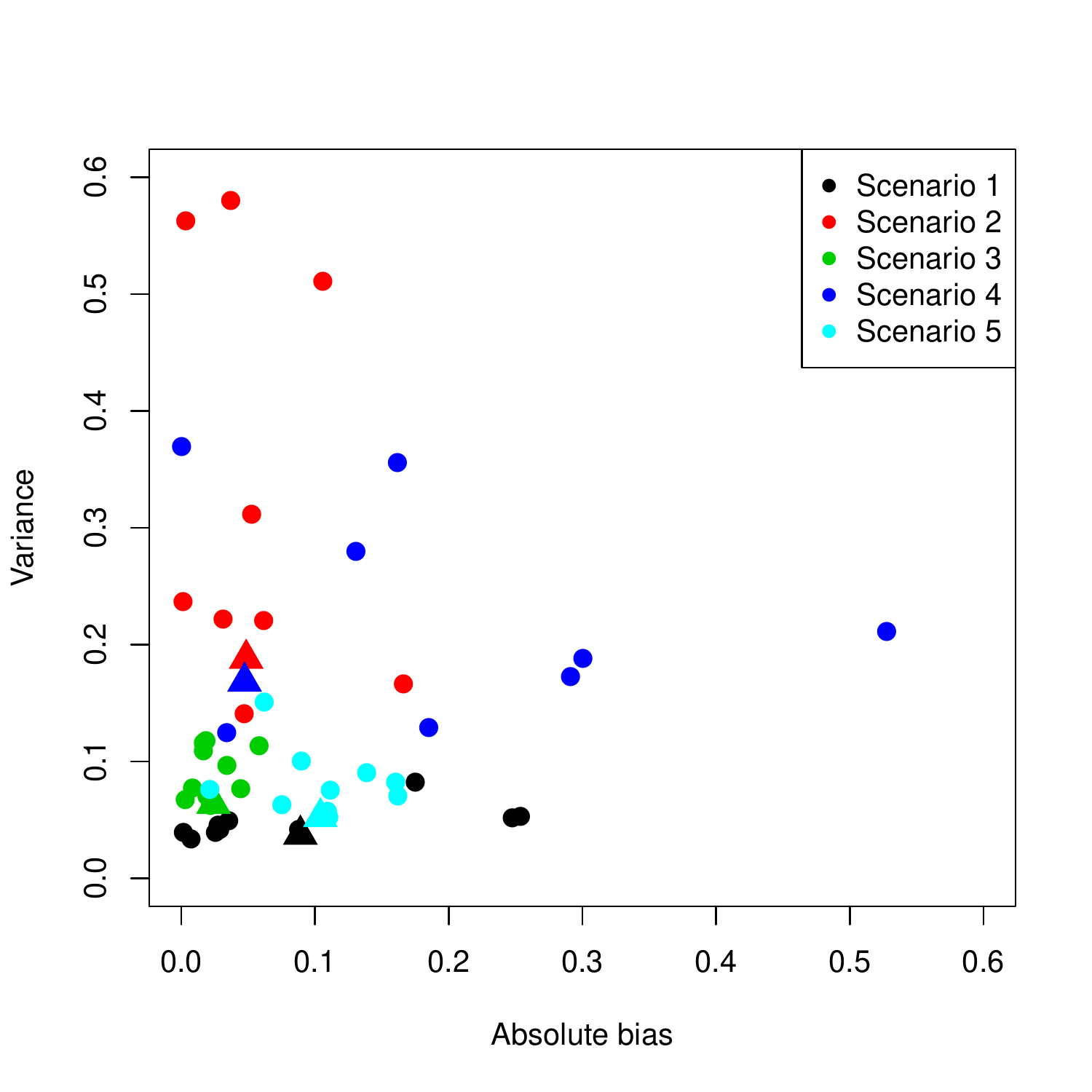}
\caption{Absolute bias and variance for each of the estimators, over all scenarios looked at. The averaged estimator is denoted by large triangles, while the other estimators are denoted by small circles.}
\label{fig:simAll}
\end{figure}

To assess the impact of bias and variance of the averaged estimator, Figure \ref{fig:simAll} shows the absolute bias and variance for each individual estimator across all the simulation scenarios considered. The circles represent the individual estimators while the triangles represent the averaged estimator for each scenario. The averaged estimator had small variance relative to the individual estimators, which is expected given the variance gains that are possible as indicated in Section \ref{sec:var}. In terms of bias, the averaged estimator is guaranteed to have exactly the average bias of all of the existing estimators. This can lead to reductions in absolute bias in cases where certain estimators have positive bias while others have negative bias, which was observed in Scenario 4. In other cases, such as Scenarios 2 and 5, the direction of the bias for the individual estimators was the same, and the absolute bias of the averaged estimator was simply the average of the absolute biases for the individual estimators. 

\section{Application to NHANES data}

Environmental wide association studies (EWAS) measure a variety of chemicals and toxins that humans are exposed to in an effort to understand how they affect the biological processes in the human body \citep{wild2005complementing,patel2014studying}. The National Health and Nutrition Examination Survey (NHANES), is a cross-sectional data source made publicly available by the Centers for Disease Control and Prevention (CDC), and we aim to estimate the effects of a number of environmental exposures on three different outcomes: HDL cholesterol levels, LDL cholesterol levels, and triglyceride levels in the NHANES study. We use data found in \cite{wilson2018model}, which contains a large number of potential confounders. Participants both filled out questionnaires regarding their health status, as well as receive clinical and laboratory tests that contain information on environmental factors such as pollutants, allergens, bacterial/viral organisms, chemical toxicants, and nutrients. In previous work \citep{patel2012systematic}, the environmental agents were separated into different groups containing similar agents that might affect similar biological pathways. We look at the effects of 22 different environmental agent groups on the three outcomes, leading to 66 different analyses. Each exposure is defined as the average exposure level across all agents within the same grouping. As most of the approaches considered only consider binary treatments, we dichotomized the continuous exposure by classifying each exposure as being above or below the median exposure level. Each of the environmental agents was measured in a different subset of the NHANES data, leading to different populations, sample sizes, and covariate dimensions for each of the 22 different exposures. Some of the data sets have very small $p/n$ ratios that do not require high-dimensional techniques, while others have $p/n$ ratios between 0.3-0.4 and require some form of dimension reduction. In the following sections we show how the averaged estimator performs in comparison with the individual estimators, and how this varies as a function of the $p/n$ ratio. 

\subsection{Agreement of statistical decisions}

\begin{figure}[H]
\centering
	  \includegraphics[width=0.7\linewidth]{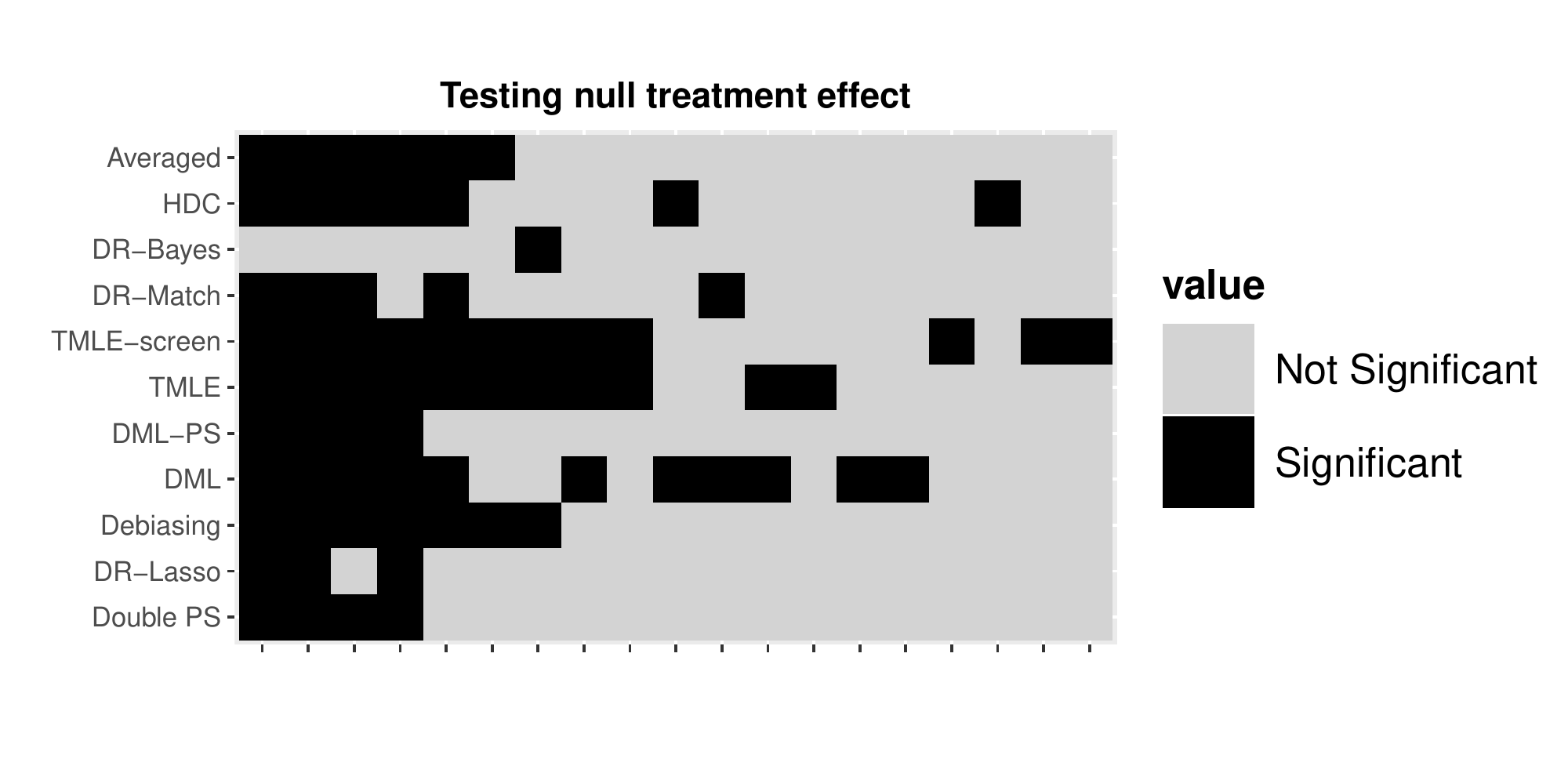}
\caption{Plot of whether or not each estimator rejected a particular null hypothesis. The results only show data sets in which there was some disagreement among the 11 estimators in terms of the final hypothesis test.}
\label{fig:Testing}
\end{figure}

First, we evaluate whether the averaged estimator would lead to different statistical inferences. In particular, we are interested in testing the null hypothesis of no treatment effect, and evaluate if the different estimators lead to different conclusions of statistical significance (i.e., is the null hypothesis rejected). Figure \ref{fig:Testing} shows the results of the hypothesis tests for any data set in which there was disagreement among the estimators. Each row corresponds to a particular estimator, while each column corresponds to a particular data set. The averaged estimator generally sides with the majority of the individual estimators in terms of rejecting or failing to reject the null hypothesis, although this is not always the case. In the sixth column, 7 of the 10 estimators failed to reject the null hypothesis of no treatment effect, but the averaged estimator leads to a rejection of the null hypothesis. It is important to emphasize that Figure \ref{fig:Testing} is limited to cases where different estimators lead to different conclusions, which represents only 30\% of analyses performed.

\subsection{When does averaging matter?}

\begin{figure}[htbp]
\centering
	  \includegraphics[width=0.45\linewidth]{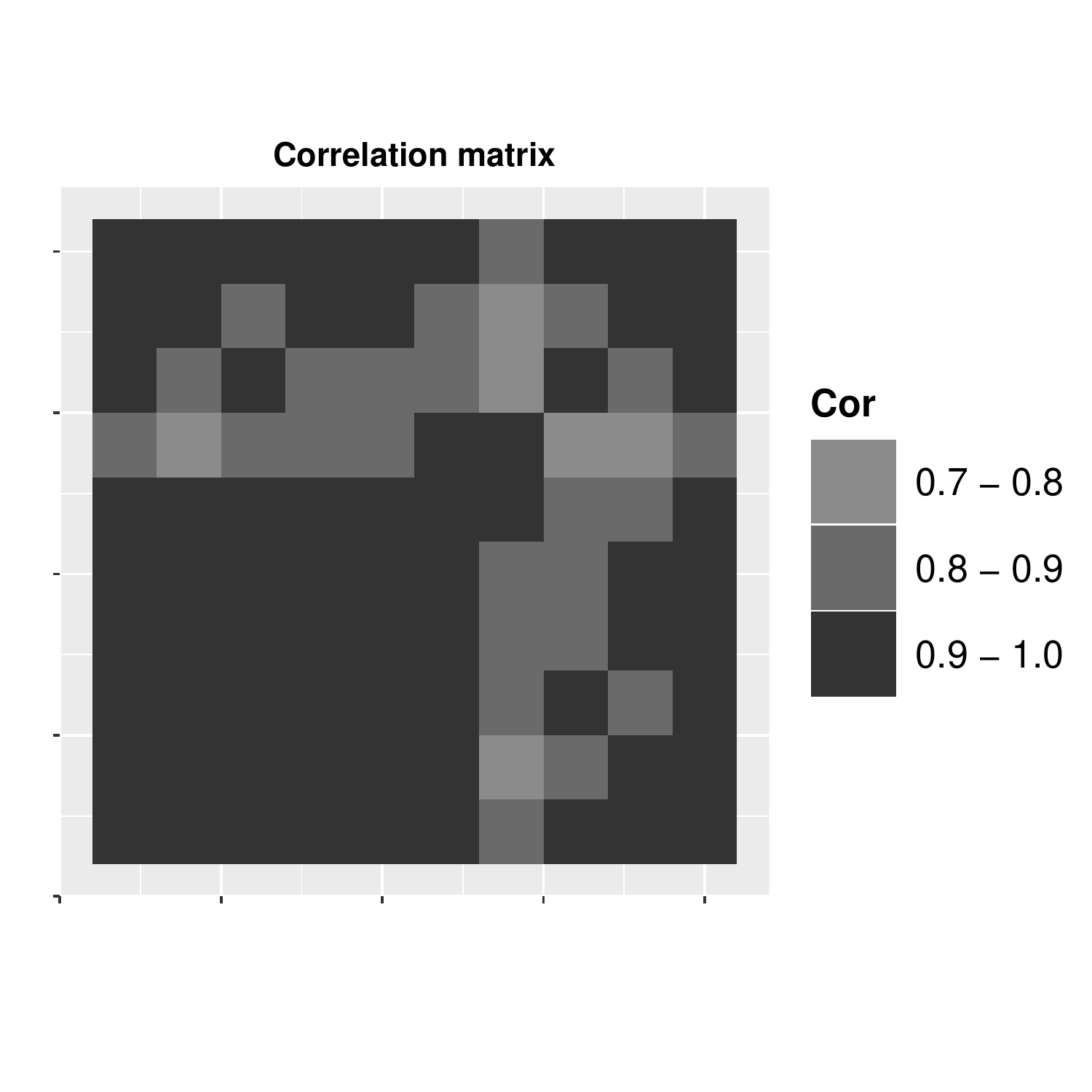}
	  \includegraphics[width=0.45\linewidth]{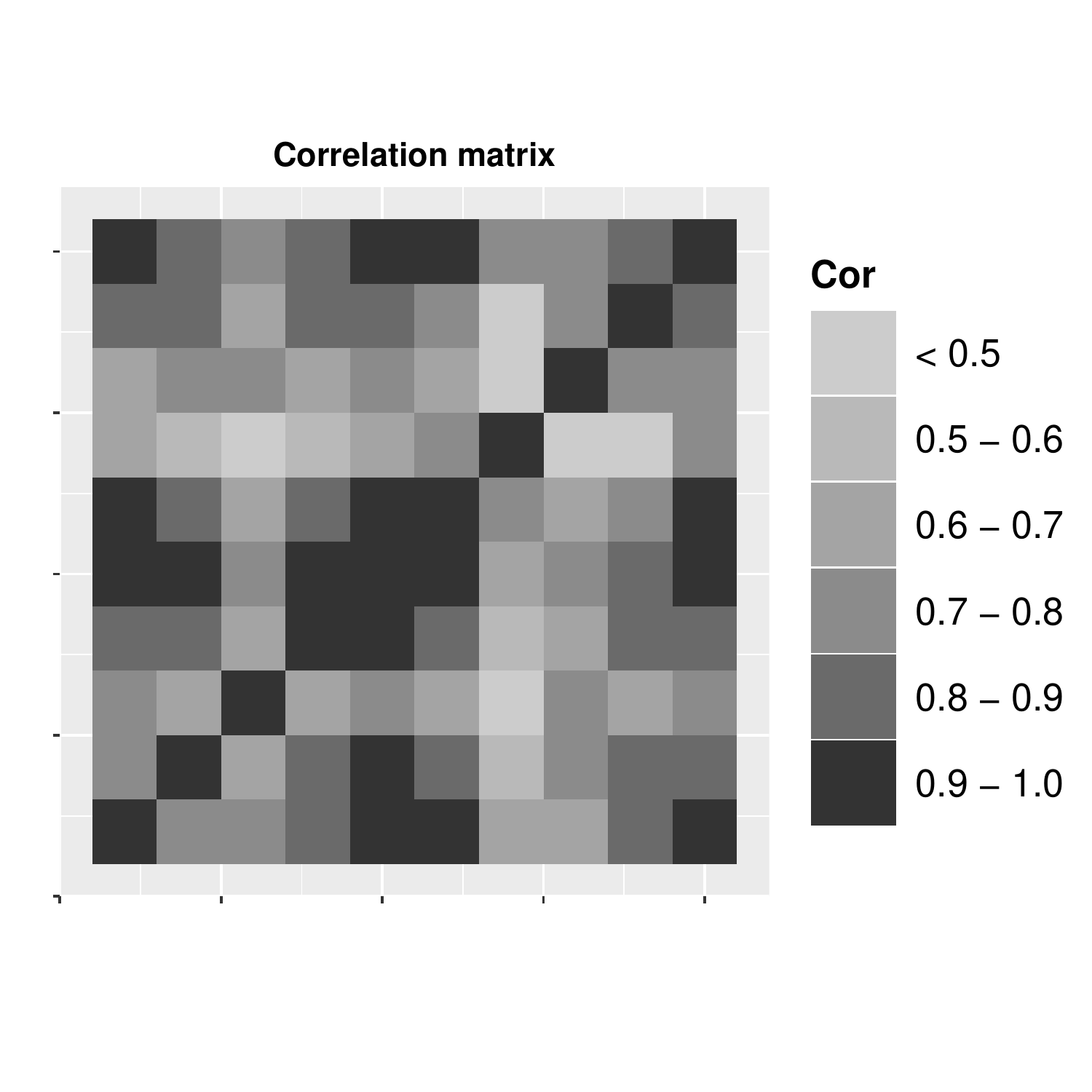}
\caption{Correlation matrix across the 10 estimators over the analyses in the NHANES data. The right panel only shows this correlation for those data sets with a $p/n$ ratio that is above 0.25}
\label{fig:CorMat}
\end{figure}

Next, we evaluated the potential impact that averaging can have on estimation within the NHANES data set, and attempted to identify where it is the most useful. Assuming that each estimator has only a small amount of bias, the main advantage of averaging is the efficiency gains one can achieve. When the individual estimators are highly correlated, there is very little to be gained (though also very little to be lost) by using averaging. On the other hand, when estimators are weakly correlated with each other, the averaged estimator gains more efficiency. The left panel of Figure \ref{fig:CorMat} shows the correlation matrix of the 10 estimators across the data sets examined. The correlation is generally very high, with the majority of the values on the off-diagonal above 0.9. The right panel, however, depicts a somewhat different story when we restricted attention to the data sets with $p/n$ ratios above 0.25. In these settings, there is less correlation between the estimators, with some of the correlations as low as 0.4, and many more in the 0.6 to 0.8 range. This illustrates that the potential for variance reduction by averaging estimators is greater in the the more difficult setting with a larger number of covariates relative to the sample size.



\section{Discussion}

We presented a simple approach to estimate the average treatment effect in high-dimensional settings that leverages many of the estimators that have been developed in recent years. As more approaches are developed, it becomes increasingly difficult for an analyst to choose an approach for their particular study. Further, there is likely no best approach in all scenarios, and the extent to which each estimator performs is highly dependent on the unknown data generating process. Averaging over all of the estimators provides an ideal choice when presented with so many estimators. Not only does the averaged estimator guarantee improved performance over the worst estimator in terms of bias and MSE, but our empirical results indicate that it can perform much more closely to the best of the individual estimators without knowing beforehand which estimators are best. At times, it can even outperform the best of the individual estimators with respect to MSE or interval coverage. Further, we derived a conservative estimate of the variance of the averaged estimator that leads to reasonable confidence interval coverage. 

A potential criticism of the averaged estimator is that each individual estimator relies on its own set of assumptions, and therefore the averaged estimator assumes all of these assumptions in order to achieve unbiased inference. While this is generally true, this in some ways shows the strength of the averaged estimator. All of these assumptions are untestable, and we don't know, and cannot test, which estimators' assumptions are being violated. By averaging over all estimators, we mitigate the impact that gross misspecification of one of these assumptions could have if we simply chose the one, bad estimator.  An analyst could alternatively choose their preferred estimator or the one with the minimum estimated variance among all candidate estimators. While this may work in certain scenarios, it can perform extremely poorly if the chosen estimator has a high bias. Further, simply choosing the estimator with the smallest estimated variance will likely lead to anti-conservative confidence intervals. The averaged estimator removes the burden of this decision from the analyst and protects against an estimator with high bias, high variance, or low interval coverage. We believe that including as many estimators as possible into the list of candidate estimators is ideal unless some estimators are known a priori to have high bias or variance. Including more estimators reduces the impact that a small number of bad estimators can have, while increasing the efficiency of the averaged estimator. Empirically we explored the possibility of using a trimmed average estimator that removes extreme estimators from the candidate list. Trimmed estimators such as this can fix the issue of one or two estimators having extremely large bias or variance, and can still lead to improvements in efficiency and interval coverage of the estimators. 

\section*{Software}

An R package for implementing the averaging estimator can be found at \\ \url{https://github.com/jantonelli111/AveragingCausalHD}.

\bibliographystyle{authordate1}
\bibliography{HDA}

\appendix

\section{Proof of result 1}

First we will show the result for bias:

\begin{align*}
    \bigg| E \bigg(\frac{1}{M} \sum_{m=1}^M \widehat{\theta}_m \bigg) - \theta \bigg| &= \bigg| \frac{1}{M} \sum_{m=1}^M \big( E(\widehat{\theta}_m) - \theta \big) \bigg| \\
    &\leq \frac{1}{M} \sum_{m=1}^M \big| E(\widehat{\theta}_m) - \theta \big| \\
    &\leq \frac{1}{M} \sum_{m=1}^M \big| E(\widehat{\theta}^*) - \theta \big| \\
    &= \big| E(\widehat{\theta}^*) - \theta \big|
\end{align*}

\noindent Now for the MSE:

\begin{align*}
    E \bigg[ \bigg( \frac{1}{M} \sum_{m=1}^M \widehat{\theta}_m  - \theta \bigg)^2 \bigg] &= E \bigg[ \frac{1}{M^2} \bigg( \sum_{m=1}^M (\widehat{\theta}_m  - \theta) \bigg)^2 \bigg] \\
    &= \frac{1}{M^2} E \bigg[ \sum_{m=1}^M (\widehat{\theta}_m  - \theta)^2 + \sum_{i \neq j} (\widehat{\theta}_i - \theta)(\widehat{\theta}_j - \theta) \bigg] \\
    & \leq \frac{1}{M^2} E \bigg[ \sum_{m=1}^M (\widehat{\theta}_m  - \theta)^2 + \sum_{i \neq j} |\widehat{\theta}_i - \theta||\widehat{\theta}_j - \theta| \bigg] \\
    & \leq \frac{1}{M^2} \bigg[ M * E\big((\widehat{\theta}^{**}  - \theta)^2 \big) + \sum_{i \neq j} \sqrt{E\big((\widehat{\theta}_i  - \theta)^2 \big) E\big((\widehat{\theta}_j  - \theta)^2 \big)} \bigg] \\
    & \leq \frac{1}{M^2} \bigg[ M * E\big((\widehat{\theta}^{**}  - \theta)^2 \big) + \sum_{i \neq j} E\big((\widehat{\theta}^{**}  - \theta)^2 \big) \bigg] \\
    &= E\big((\widehat{\theta}^{**}  - \theta)^2 \big)
\end{align*}

\section{Computational details for each estimator}

\begin{figure}[H]
\centering
	  \includegraphics[width=0.45\linewidth]{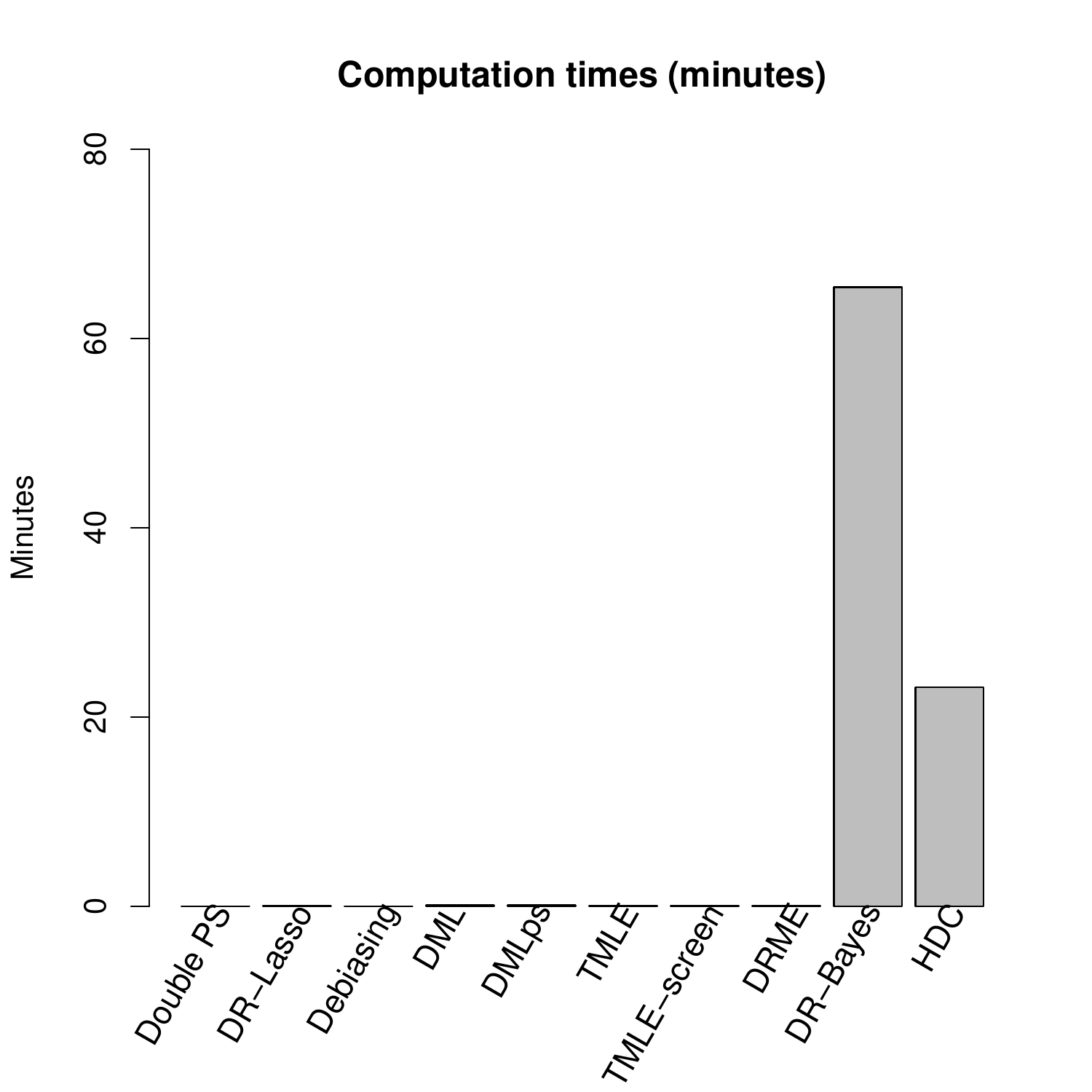}
\caption{Computation time for individual estimators}
\label{fig:CompTimes}
\end{figure}

Here we illustrate the computation time of the 10 individual estimators used in the manuscript when dimension of the data is $n=150$ and $p=300$. The computation time (in minutes) can be found in Figure \ref{fig:CompTimes}. All of the approaches have computation times smaller than 10 seconds with the exception of DR-Bayes and HDC, which is unsurprising given that both of these approaches require MCMC chains to be run for high-dimensional Bayesian models. If computational resources and time are a priority, then the averaged estimator can be fit without including these Bayesian approaches into the library of estimators. 

\section{Data generating mechanisms for simulation scenarios and additional simulation results}

Here we will detail the exact data generating processes for each of the simulation scenarios considered in the paper. In addition, we will provide five new simulation scenarios and present the results for them as well. 

\textit{Scenario 1:} For the first simulation scenario, we generated data under the following models for the treatment and the outcome processes:

\begin{align}
    Y_i &= T_i + \boldsymbol{X}_i \boldsymbol{\beta} + \epsilon_i \label{eqn:sim} \\
    P(T_i = 1) &= \Phi(\boldsymbol{X}_i \boldsymbol{\gamma}) \nonumber \\
    \boldsymbol{X}_i &\sim \text{MVN}(\boldsymbol{0}_p, \boldsymbol{\Sigma}_X) \nonumber \\
    \epsilon_i &\sim \mathcal{N}(0,1), \nonumber
\end{align}

\noindent where $\boldsymbol{\Sigma}_X$ is an exchangeable correlation matrix with correlation 0.3, $n=150$, $p=300$, $$\boldsymbol{\beta} = (0.75,1, 0.6, -0.8, -0.7, \boldsymbol{0}_{p-5}),$$ and $$\boldsymbol{\gamma} = (0.15,0.2,0, 0, -0.4, \boldsymbol{0}_{p-5}).$$

\textit{Scenario 2:} This simulation is based on a simulation from \cite{athey2018approximate} that has a dense treatment model. Define 20 clusters, $\{\mathbf{c_1},\dots, \mathbf{c_{20}} \}$ where $\mathbf{c_k} \sim \mathcal{N}(0, I_{p})$. We draw $\mathbf{C}_i$ uniformly at random from one of the 20. Then, we draw the covariates from a multivariate normal distribution centered at $\mathbf{C}_i$ with the identity matrix as the covariance. We set $T_i = 1$ with probability 0.1 for the first 10 clusters, and $T_i = 1$ with probability 0.9 for the remaining clusters. Finally, we generate data from the outcome model defined as $Y_i = 10 T_i + \boldsymbol{X \beta} + \epsilon_i$, where $\boldsymbol{\beta} \propto (1, \frac{1}{\sqrt{2}}, \dots, \frac{1}{\sqrt{p}})$ and is normalized such that $|| \boldsymbol{\beta}||_2^2 = 18$. Lastly, $n=300$ and $p=300$.

\textit{Scenario 3:} This scenario uses the same structure as in equation \ref{eqn:sim} and simulation scenario 1, with the exception of $\boldsymbol{\beta}$ and $\boldsymbol{\gamma}$. In this case, we set $\boldsymbol{\beta} = (0.2, 0.3, 0.15, -0.15, 0.15, -0.15, \dots 0.15, -0.15, \boldsymbol{0}_{p-74})$, and $\boldsymbol{\gamma} = (0.6,-0.5, \boldsymbol{0}_{p-2})$.

\textit{Scenario 4:} This scenario uses the same simulation structure as scenario 3, except the outcome model is chosen to be nonlinear. Now we generate the outcomes from a normal distribution with variance 1 and mean given by $$E(Y_i \vert T_i, \boldsymbol{X}_i) = -2 + T_i + 0.5 X_{1i} + 0.5 X_{2i}^2 + 0.8 X_{2i}^3 + 0.3 e^{X_{3i}}$$

\textit{Scenario 5:} This scenario uses the same simulation structure as simulation 1 except for the treatment and outcome models. In this setting, we have $P(T_i = 1) = \Phi(-0.7 + \boldsymbol{X \gamma})$, where $\boldsymbol{\gamma} = (0.2, 0,0.4, -0.5, \boldsymbol{0}_{p-4})$ and we generate the outcome with a variance of 1 and mean given by $$E(Y_i \vert T_i, \boldsymbol{X}_i) = -2 + T_i + 0.5 X_{1i} + 0.8 X_{2i} + 0.4 X_{3i} + 0.5 T_i X_{3i}$$

\textit{Additional scenario 1:} This scenario uses the same simulation structure as the first simulation scenario of the main manuscript, however, both $\boldsymbol{\beta}$ and $\boldsymbol{\gamma}$ are $\boldsymbol{0}_p$ and there is no confounding.

\textit{Additional scenario 2:} This scenario uses the same simulation structure as the first simulation scenario of the main manuscript, however, $$\boldsymbol{\beta} = (0.9, 0.9, 0.2, 0.2, 0, 0, 0.9, 0.9, \boldsymbol{0}_{p-8}),$$ and $$\boldsymbol{\gamma} = (0.4, 0.9, -0.5, -0.7, -0.3, 0.6, \boldsymbol{0}_{p-6}).$$

\textit{Additional scenario 3:} This scenario is the same as the second simulation scenario of the main manuscript, except that $T_i=1$ with probability 0.25 and 0.75 for the first 10 and remaining clusters, respectively. Further, we set  $\boldsymbol{\beta} \propto (1, \frac{1}{2}, \dots, \frac{1}{p})$.

\textit{Additional scenario 4:} This scenario is the same as the first simulation scenario of the main manuscript except the treatment model is nonlinear. In particular the treatment probabilities are given by $$P(T_i = 1) = \text{expit}(0.5X_{1i}^3 + 0.3X_{1i}^2 - 0.3X_{2i}^4 + 0.4*X_{3i}^2)$$

\textit{Additional scenario 5:} This scenario is the same as the fourth additional simulation scenario except that the outcome model is also nonlinear. Specifically, the outcome is drawn from a normal distribution with variance 1 and mean given by $$E(Y_i \vert T_i, \boldsymbol{X}_i) = -2 + T_i + 0.5 X_{1i} + 0.5 X_{2i}^2 + 0.8 X_{2i}^3 + 0.3 e^{X_{3i}}$$

\begin{figure}[H]
\centering
	  \includegraphics[width=0.9\linewidth]{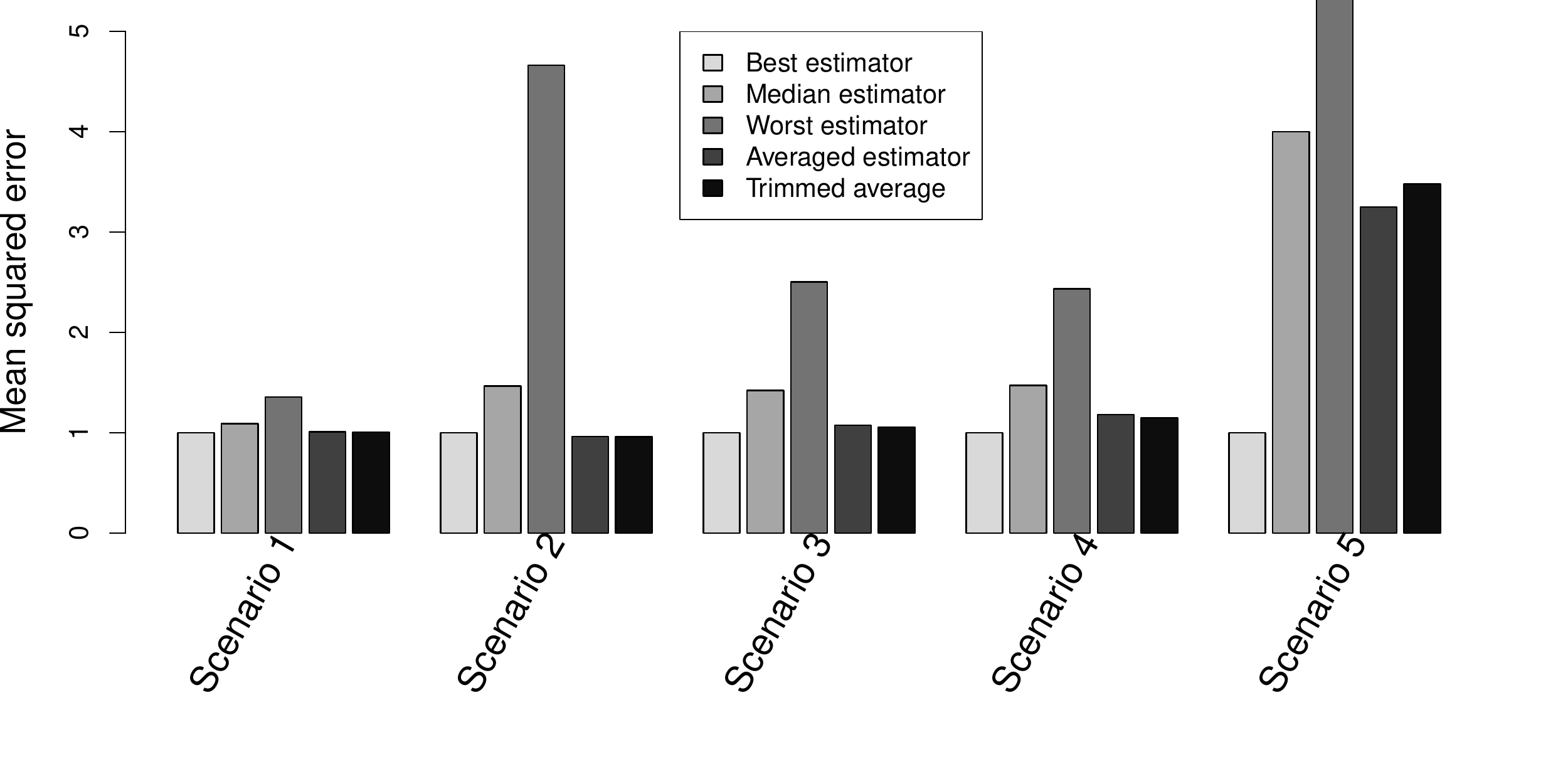}
\caption{Mean squared error for the estimator with the lowest MSE, highest MSE, median MSE, the averaged estimator, and the trimmed average estimator. Note that these results are for the five additional simulation results of the appendix.}
\label{fig:simMSEapp}
\end{figure}

\begin{figure}[H]
\centering
	  \includegraphics[width=0.9\linewidth]{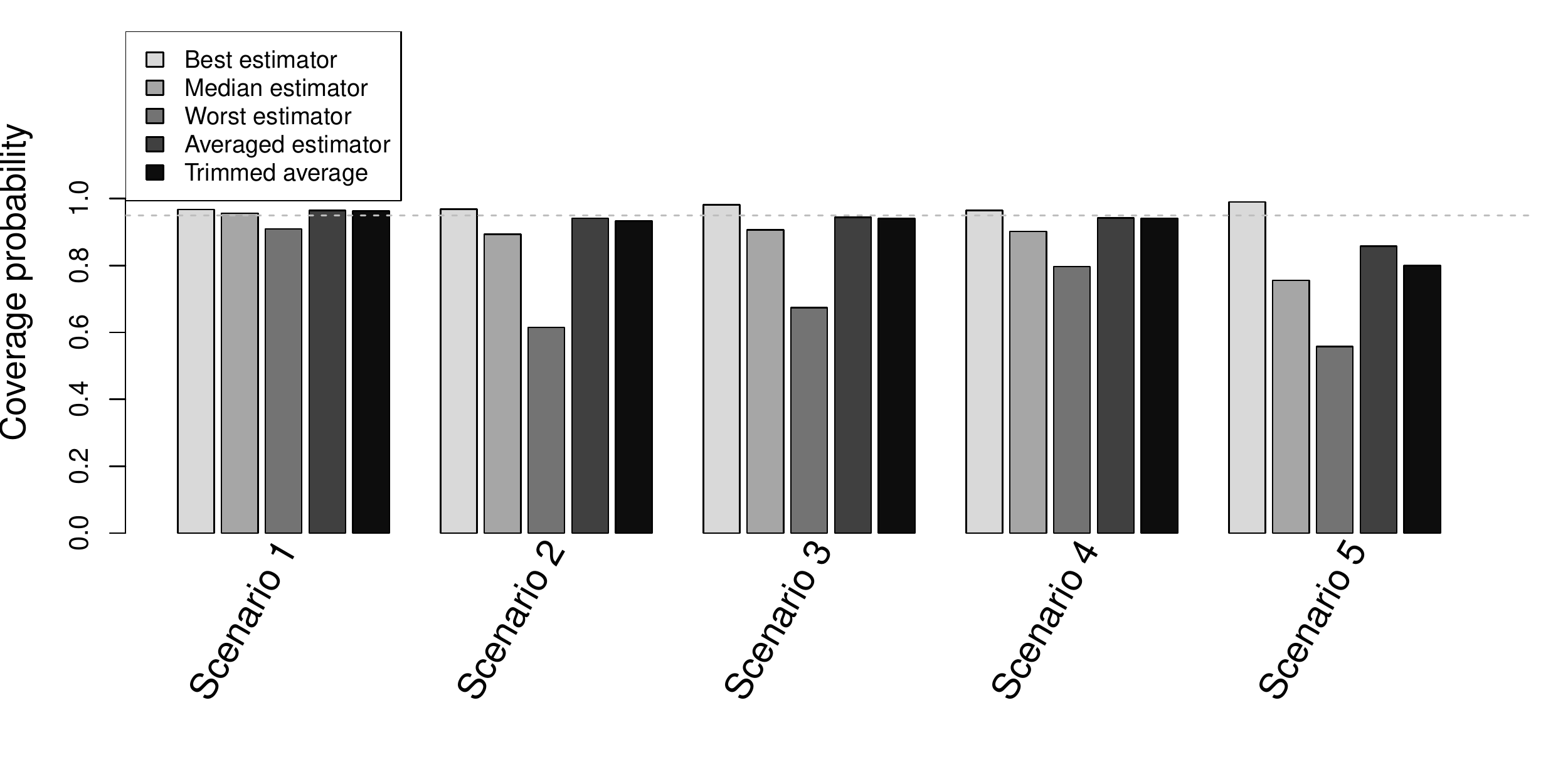}
\caption{Interval coverage for the estimator with the best interval coverage, median interval coverage, worst interval coverage, the averaged estimator, and for the trimmed average estimator. Note that these results are for the five additional simulation results of the appendix.}
\label{fig:simINTapp}
\end{figure}

The first additional simulation scenario looks at a scenario with no confounding among the covariates. The second additional scenario looks at a different sparse data generating process for both the treatment and outcome models. The third additional scenario looks at a different scenario where the treatment model is not sparse. The fourth scenario looks at a case where the treatment model is nonlinear and therefore most approaches will be misspecified. The fifth additional scenario looks at a case where both the treatment and outcome models are nonlinear. The results from these additional simulations can be found in Figures \ref{fig:simMSEapp} and \ref{fig:simINTapp}. In every one of the scenarios looked at the averaged estimators do far better than the worst estimator and better than the median estimator. In some cases, the averaged estimators do far better than the median estimator and perform as well or better than the best individual estimator. The averaged estimator also obtains the nominal 95\% coverage rate in every simulation scenario except for the fifth scenario. The only scenario in which the averaged estimators were substantially outperformed by the best individual estimator was scenario 5. In this scenario there was only one estimator that was unbiased and had nominal confidence interval coverages. The remaining estimators all had substantial biases and low interval coverages. Overall these results confirm the results from the manuscript. The averaged estimator, whether it is trimmed or not, shows performance that is much closer to the best individual estimator than either the median or worst individual estimator. This leads to low MSE values and interval coverages that are much closer to the nominal rate when compared with the majority of the individual estimators. 

\end{document}